\documentclass[preprint]{aastex631}

\begin{document}

\title{Galaxy dynamics tracing quantum cosmology beyond $\Lambda$CDM below the de Sitter scale of acceleration}

\correspondingauthor{Maurice H.P.M. van Putten}
\email{mvp@sejong.ac.kr}
\author[0000-0002-0786-7307]{Maurice H.P.M. van Putten}
\affiliation{Physics and Astronomy, Sejong University\\
209 Neungdong-ro, Seoul, South Korea }

\affiliation{INAF-OAS Bologna, via P. Gobetti, 101, I-40129 Bologna, Italy}

\begin{abstract}
It is proposed that the baryonic Tully-Fisher relation (bTFR) and JWST 'Impossible galaxies' at cosmic dawn are unified in weak gravity by the trace $J$ of the Schouten tensor below the Sitter scale of acceleration $a_{dS}=cH$, where $c$ is the velocity of light and $H$ is the Hubble parameter. Across $a_{dS}$, $J$ parametrizes short-period galaxy rotation curves and fast gravitational collapse beyond the predictions of $\Lambda$CDM.
The sensitivity of weak gravitation to $J=\frac{1}{6}R$ is derived from in infrared gravitation from a consistent limit of quantum gravity, reducing to general relativity in the limit of a small Planck constant, where $R$ is Ricci scalar tensor.
For the first time, it identifies the exact relation 
$a_0=c^2\sqrt{J}/2\pi$ of the Milgrom parameter across all redshifts, accounting for the bTFR and early galaxy formation accelerated by a factor $\sim J^{1/8}$. It predicts $a_0^\prime(0)<0$ at the present epoch.    
\end{abstract}

\keywords{Galaxy dynamics -- cosmology}

\section{Introduction}

The Universe largely evolves by gravitation, from structure formation structure at cosmic dawn \citep{eis23} to black hole 
formation \citep{gen03,che03,che08,gen10,abb16,EHT22}. 
On the largest scales, its evolution satisfies the Friedmann-Lema\^itre-Robertson-Walker (FLRW) line-element
\begin{eqnarray}
    ds^2 = a^2\eta_{ab}dx^a dx^b,
    \label{EQN_01}
\end{eqnarray}
describing the conformal scaling of Minkowski spacetime $\eta_{ab}$ by the Friedmann scale $a$. The Hubble rate of expansion $H=\dot{a}/a$ is given in the equivalent line-element $ds^2=-c^2 dt^2 + a(t)^2\left(dx^2+dy^2+dz^2\right)$ of a Big Bang cosmology with a singularity at a Hubble time $1/H_0$ in the past \citep[e.g.][]{haw70,oli22}, where $c$ is the velocity of light.
Presently, expansion is accelerating evidenced by a deceleration parameter $q_0<0$ \citep{per99,rie98,rie22}, 
where $q(t)=-a\ddot{a}/\dot{a}^2$, equivalently, $q(z)=-1+(1+z)H^{-1}(z)H^\prime(z)$ by cosmological redshift $z$ in $a/a_0=1/(1+z)$.

On sub-horizon scales below the Hubble radius $R_H=c/H$, the large-scale structure forms by gravitation, largely in weak gravitation
at or below the de Sitter scale of acceleration of the cosmological background (\ref{EQN_01}) \citep{van17a,van17b,van17c}
\begin{eqnarray}
    a_{dS}=cH \simeq 7{\rm \AA} \,{\rm s}^{-2},
    \label{EQN_adS}
\end{eqnarray}
where the second equality refers to the present epoch.
Crucially, weak gravitation reveals anomalously short timescales in galaxy dynamics, described by the baryonic Tully-Fisher relation (bTFR) \citep{tul77,mcc12}, and, recently, in
galaxy formation at cosmic dawn inferred from the JWST `Impossible galaxies' \citep[e.g.][]{boy24}.
In $\Lambda$CDM, the bTFR is attributed to dark matter on galactic scales. Alternatively, it may derive from a finite sensitivity of radial acceleration in weak gravitation to $a_{dS}$ in a departure from Newton's second law of motion. 
If so, this points to a possible unification with early galaxy formation by fast gravitational collapse at cosmic dawn, when dark matter is already saturated essentially at closure density for the latter \citep{van24d}, upending a key assumption 
of $\Lambda$CDM on sub-horizon scales.

To be sure, the Hamiltonian of classical mechanics is invariant under a change of inertia. For circular motion in galaxy rotation curves, centripetal accelerations $\alpha = V_c^2/r$ in orbital motion at circular velocity $V_c$ and radius $r$ based on spectroscopic observations commonly exceed the expected Newtonian acceleration: $\alpha > V_N^2/r=a_N$, where $a_N=GM_b/r^2$ derives from baryonic mass $M_b$ in stars and gas given
Newton's constant $G$. Conserving momentum, $\alpha m^\prime = F_N$ defines non-Newtonian inertia $m^\prime$ subject to the Newtonian acceleration $F_N=ma_N$, given the gravitational binding energy $U_N=-\int_r^\infty F_N ds = -GmM_b/r$ to $M_b$.
For a star of mass $m$, the Newtonian kinetic energy $E_k=\frac{1}{2}mV_N^2$, $a_N=V_N^2/r$, and the observed kinetic energy $E_k^\prime=\frac{1}{2}m^\prime V_c^2$ satisfy
\begin{eqnarray}
E_k^\prime = \frac{1}{2} m V_c^{ 2} \left( \frac{m^\prime}{m}\right) =  \frac{1}{2} m V_c^{2} \left(\frac{a_N}{\alpha}\right) 
= \frac{1}{2} m V_N^2.
\end{eqnarray}
The Hamiltonian $H = E_k + U_N$ is hereby invariant under a change of inertia to $m^\prime$ - a symmetry of classical mechanics that allows a finite sensitivity of galaxy dynamics to $a_{dS}$ with no need for dark matter on galactic scales nor variation of fundamental constants \citep{van24a,van24b}.

We here approach a sensitivity of $m^\prime$ to $a_{dS}$ based on the quantum mechanical properties of particles by {\em position} according to Compton phase $\varphi$. Since general relativity
is parameterized by $G,c$ but not $\hbar$, consistent coupling
to gravitation requires regularization in light of UV-divergence $\varphi \sim 1/\hbar$. 
(Similar considerations apply to the bare cosmological constant $\Lambda_0\sim 1/\hbar$ \citep{wei89,van24c}.)

A {\em primitive} as such, $\varphi$ requires completion in an IR consistent coupling $\sim \hbar$ to general relativity, 
here considered by the IR/UV-relation \citep{van24c}
\begin{eqnarray}
    \alpha_p A_p=1
    \label{EQN_IR}
\end{eqnarray}
with $A_p$ the area of concentric 2-spheres in units of Planck area $l_p^2$, $l_p=\sqrt{G\hbar/c^3}$.

To quantify the finite sensitivity of weak gravitation to $a_{dS}$, we revisit particle motion in the Newtonian limit (\S2) and IR consistent coupling giving rise to curvature (\S3).
The cosmological vacuum is seen to have a finite temperature associated with $a_{dS}$. The Unruh temperature seen by accelerating particles hereby scales with the index of refraction of the vacuum, deviating from unity in weak gravitation below $a_{dS}$ (\S4).
Accordingly, inertial back reaction picks up dispersion in weak gravitation at accelerations below $a_{dS}$ (\S5). 
These states of motion near free fall are currently
outside the realm of laboratory experiments, where total accelerations are subject to the Earth's 
gravitational field \citep{gru07}, yet are naturally present 
in radial accelerations in galaxy dynamics and galaxy formation.
For these cases, we derive asymptotic results allowing for a confrontation with data from spiral galaxies and galaxy formation at cosmic dawn (\S6). 

\section{Equivalence principle in the Newtonian limit}

By the equivalence principle, a particle at acceleration $\kappa$ 
acquires a potential energy $U>0$ in the gravitational field $g=-\kappa$.
For a particle of mass energy $E=mc^2$ on a flat spacetime background, $U$ relative to the Rindler horizon $h$ of its comoving frame satisfies
\begin{eqnarray}
    U=\int_0^\xi F_\kappa ds = E,
    \label{EQN_U}
\end{eqnarray}
where $\xi=c^2/\kappa$ is the distance to $h$ given the velocity of light $c$. 
In the equivalent work against $F_\kappa = \kappa m$ across $\xi$, (\ref{EQN_U}) shows heat $Q=U$ extracted from $h$.
It represents a Clausius integral $U=\int_0^m T_\kappa dS$ at  thermodynamic temperature, 
\begin{eqnarray}
 k_BT_\kappa=  \left(\frac{\partial S}{\partial E}\right)^{-1} = \frac{\kappa\hbar}{2\pi c},
 \label{EQN_TU}
\end{eqnarray}
where $k_B$ is the Boltzmann constant, i.e., the Unruh temperature of $\kappa$ with entropy $S=k_BI$ by Compton phase over $\xi$,
\begin{eqnarray}
    I=2\pi\varphi
    \label{EQN_I}
\end{eqnarray}
relative to $h$ based on the particle's Compton wave number $k_C = mc/\hbar$ and propagator \citep{van15a}.

The position information (\ref{EQN_I}) is well-known from the original identification of black hole entropy by \citet{bek73}. 
It formally derives from the particle propagator. 
A mass $m$ at the Zitter angular frequency $\omega \hbar =mc^2$ in a cube of dimension $L$ requires encoding by $I=12 \varphi k_B$ based on $\varphi=\omega s/c$ over $s=L/2$ \citep{van15a}.
For a concentric 2-sphere of radius $s$, $I$ is reduced by orthogonal projection of the area $24s^2$ 
of the cube to the area $4\pi s^2$ by the area ratio $\pi/6$. 
It leaves $I = 2\pi \varphi$ universally, also for $\epsilon_\nu=\omega\hbar$ of radiation at the 
Unruh temperature $k_BT_h=\left(\partial S/\partial \epsilon\right) = \hbar c/2\pi\xi$ of the Rindler horizon $h$ at distance $\xi=c^2/\kappa$ at acceleration $\kappa$. 
The Boltzmann factor $e^{-\hbar\omega/k_BT}$ of 
Unruh radiation has a corresponding phase jump $\varphi = \omega\xi/c$, whereby ${\hbar\omega}/{k_BT}=2\pi \varphi$ once more. (With zero rest mass, however, $\varphi$ of (Unruh) radiation is not divergent in $1/\hbar$.) It highlights $I$ to be an energy ratio which, for a massive particle, derives from its Zitter frequency. 
By $e^{-\hbar\omega/k_BT}=e^{-2\pi\varphi}$, $I$ is recognized to be a primitive, here with entropy $S=k_BI$ in the Boltzmann factor of $h$. As such, $I$ is a {\em bare massless scalar field} in light of the UV-divergence $I\sim 1/\hbar$. 

Crucially, for massive particles, $I$ is a primitive by its UV-divergence $I\sim 1/\hbar$, that requires an IR consistent completion in $U$ by coupling aforementioned massless modes at $k_BT_\kappa\sim \hbar$ to gravitation. 
Including Newton's constant, $G$, $I$ has a counterpart in spacetime curvature as follows.

\section{Spacetime curvature in infrared gravitation}

About a gravitating mass $M$, a natural continuation to 2-surfaces beyond the Schwarzschild radius $R_S=2R_g$, $R_g=GM/c^2$, results in the Bekenstein bound \citep{bek81} on the phase space within, i.e., $I\le S/k_B$. 
The primitive (\ref{EQN_I}) is hereby accompanied by the {\em Einstein area} $A_E=4Il_p^2$ in concentric 2-spheres of area $4\pi r^2$ at radius $r\ge R_S$. The factor four derives from the entropy $S=k_BI$ relative to the Rindler horizon $h$ of \S2, identified with the Bekenstein-Hawking entropy by application of the equivalence principle \citep{van24d}.

Specifically, $A_E$ represents a fraction of the area of wavefronts of constant phase $A_\varphi$ 
(see also \cite{van12}) at density 
\begin{eqnarray}
p= \frac{A_E}{4\pi r^2} = 4I\alpha_p.
\label{EQN_p1}
\end{eqnarray}
Conversely, the Newtonian potential (\ref{EQN_u}) is a fraction $p$ of the area $A_p$ in IR consistent coupling of $I$ to gravitation. A remainder $1-p$ is free, allowing $A_\varphi$ to propagate (by Huygens principle) at velocity $\beta=v/c$,
\begin{eqnarray}
 \beta = 1- p.
 \label{EQN_p}
\end{eqnarray}
The result is commonly expressed as gravitational redshift 
$\sqrt{\beta}$ in the spherically symmetric 
line-element $ds^2=\beta c^2dt^2 + dr^2/\beta$ in $(t,r)$.
Null-geodesics satisfy the {\em regressed} propagation velocity $\beta = dr/cdt$, in spacetime curvature by consistent IR coupling of (\ref{EQN_I}).
The Newtonian limit $\sqrt{\beta}\simeq 1+u$, shows a completion of $I$ in the dimensionless Newtonian binding energy
\begin{eqnarray}
    u=-2I\alpha_p = -\frac{R_g}{r}
    \label{EQN_u}
\end{eqnarray}
in a consistent IR coupling by $\alpha_p$. 
It represents the geometric embedding of the
thermodynamic temperature accompanying $I$ of \S2.
$I$ hereby finds complementary completions in temperature (\ref{EQN_U}) by $\kappa = c^2u^\prime$ and
in curvature $u^\prime\sim 1/r^2$ about a gravitating mass $M$.

Regressed wave propagation (\ref{EQN_p}) can also result from dispersion on a cosmological background with finite dark energy. 

\section{The low-energy cosmological vacuum}

On a cosmological background with dark energy, 
wave motion at low energies is dispersive. 
In weak gravitation, (\ref{EQN_U}) is hereby sensitive to $a_{dS}$ and the associated temperature of the cosmological vacuum. 

By (\ref{EQN_IR}), an IR consistent coupling of gravitation to the UV-divergent bare cosmological constant $\Lambda_0\sim 1/\hbar$ \citep{wei89} results in a dark energy $\Lambda=J$, where $J$ denotes the trace of the Schouten tensor \citep{van24c}. It derives from gauging the total phase in the path integral formulation of quantum cosmology, first introduced in \cite{van20}. 

Accordingly, the vacuum of (\ref{EQN_01}) assumes a temperature,
with reference to aforementioned deceleration parameter $q=q(z)$,
\begin{eqnarray}
T_H = \left(\frac{1-q}{2}\right) T_{dS}
\label{EQN_TH}
\end{eqnarray}
according to the surface gravity of the Hubble horizon,
where $T_{dS}$ is the de Sitter temperature associated with $a_{dS}$ \citep{gib77}.
The horizon temperature (\ref{EQN_TH}) applies to a three-flat universe (\ref{EQN_01}).
It has a dark energy given $J=\frac{1}{6}R$, defined by the Ricci
scalar-tensor $R$ of four-dimensional spacetime. 
In (\ref{EQN_01}), we have \citep{van21,van24c}
\begin{eqnarray}
    J=\left(1-q\right)H^2/c^2.
\label{EQN_J}
\end{eqnarray}
We note that $\sqrt{J}R_H = \sqrt{1-q}$.

At finite acceleration, the total vacuum temperature observed is $T=\sqrt{T_H^2+T_\kappa^2}$ based on the norm of total acceleration \citep{nar96,des97,jac98,kli11}.
On this background, acceleration $\kappa$ below $a_{dS}$ is sensitive to $J$ \citep{van17c}.
The minimal extension of (\ref{EQN_p1}) is the massless scalar field $p\rightarrow pe^{i\varphi}$. 
As a fraction of area, it satisfies the conformally invariant Klein-Gordon equation that, 
in four-dimensional spacetime, carries a coupling to $J$ \citep{wal84}. Equivalently, it satisfies coupling to the dark energy of the cosmological vacuum. 
Propagation of wavefronts $A_\varphi$ is hereby dispersive, satisfying
\begin{eqnarray}
\omega = c\sqrt{k^2 + J}
\label{EQN_disp1}
\end{eqnarray}
between angular frequency $\omega$ and wave number $k$.
According to (\ref{EQN_disp1}), the cosmological vacuum has a non-trivial refractive index $n$ for radiation at wavelengths on the order of the Hubble scale. 
%refractive index $n=c/v_g$, where $v_g=d\omega/dk = c^2k/\omega$ is the group velocity. 
The Unruh temperature $k_BT=\hbar/P$ derives in Euclidean space from the period $P=2\pi c/\kappa$ of $\theta=i\tau$, following the transformation of the line-element $ds^2 = -c^2dt^2 + dx^2 = \xi^2 d\theta^2 + d\xi^2$ to those of accelerating observers $\left(ct,x\right)=\xi\left(\sinh\lambda,\cosh\lambda\right)$,
where $\lambda = \kappa \tau/c$. It satisfies $k_BT\sim 1/c$.
While $I$ is an energy ratio (\S2), fall back to phase differences $k\xi$ in the equivalent expression of massless modes would call for an accompanying scaling of the Unruh temperature with refractive index $n^{-1} = \omega/kc$,
\begin{eqnarray}
 T_\kappa^\prime = n^{-1} T_\kappa.
 \label{EQN_Tn}
 \end{eqnarray}

\section{Galaxy dynamics in weak gravitation}

For a galaxy of mass $M=M_{11}10^{11}M_\odot$ with gravitational radius $R_G=GM/c^2$, centripetal accelerations drop below (\ref{EQN_adS}) across the transition radius \citep{van17a}
\begin{eqnarray}
    r_t = \sqrt{R_HR_G}= 4.7\,M_{11}^{1/2}{\rm kpc},
    \label{EQN_rt}
\end{eqnarray}
where the right side refers to the current epoch $z=0$. 
This transition can be seen in a normalized plot of the radial acceleration of the {\em Spitzer} Photometry and Accurate Rotation Curves (SPARC) \citep{lel16} as a function of $\zeta=a_N/a_{dS}$ \citep{van18}. It bears out in a sharp and essentially $C^0$-transition over a $6\,\sigma$ gap relative to $\Lambda$CDM galaxy models \citep{kel17}.

Here, we consider (\ref{EQN_rt}) for a transition to non-classical inertia at $\zeta<1$ due to the refractive cosmological background at low energy described by (\ref{EQN_disp1}).
For a particle of mass energy $E=mc^2$, following (\ref{EQN_U}), 
inertia $m^\prime$ appears with gravitational potential energy $U^\prime=m^\prime c^2$ in the comoving frame.  
In weak gravitation, $\alpha < a_{dS}$ and $U^\prime$ is relative to the Hubble horizon ${\cal H}$ rather than $h$ of Rindler spacetime by causality. 
Effectively, $R_H$ introduces an IR cut-off $R_H < \xi$ in the 
calculation of total gravitational potential energy, leaving $U^\prime < E$ in departure of (\ref{EQN_U}) by a factor of $\sim R_H/\xi$.

At acceleration $\alpha < a_{dS}$, the total vacuum temperature satisfies $T\simeq T_H$. 
The total heat extracted from ${\cal H}$ satisfies 
$E\simeq \frac{1}{2}k_BT_HA_p = 2Nk_BT_H$ is a sum over $N$ surface modes, where $N=\frac{1}{4}A_p$ in a 2-surface about $M$. 
For a vacuum with dispersion (\ref{EQN_disp1}), 
picked up is a finite mass in the low-energy limit $\hbar\omega\simeq \hbar c\sqrt{J}$ per propagating mode, leaving
\begin{eqnarray}
U^\prime=m^\prime c^2 = N\left(\hbar \kappa/c\right)\varphi^\prime \simeq N \left(\hbar\kappa/c\right) \sqrt{J}R_H
\label{EQN_Un}
\end{eqnarray}
by the cut-off phase $\varphi^\prime = \omega \Delta t^\prime 
\simeq \left( c\sqrt{J}\right)(R_H/c)$ in each mode relative to ${\cal H}$ rather than $h$ outside. Alternatively by (\ref{EQN_Tn}), 
$U^\prime = Nk_BT_\kappa^\prime = n^{-1}Nk_BT_\kappa$
with $\varphi^\prime = kR_H$ per mode at $T_\kappa^\prime$, arriving at the same (\ref{EQN_Un}).

By conservation of momentum $m^\prime \kappa = ma_N$, a centripetal acceleration $\alpha=\kappa$ arising from radial forcing in a Newtonian gravitational field $a_N=GM/r^2$ of $M$, it follows
\begin{eqnarray}
    \frac{a_N}{\kappa} = \frac{m^\prime}{m} = \frac{U^\prime}{E}
    \simeq \sqrt{J}R_H\frac{\hbar \alpha/c}{2k_BT_H} 
    = \frac{2\pi}{\sqrt{1-q}} \left(\frac{\alpha}{a_{dS}}\right)
    \label{EQN_N2d}
\end{eqnarray}
with notable cancellation of $N$ and using the horizon temperature (\ref{EQN_TH}).

We summarize (\ref{EQN_N2d}) by the Milgrom parameter $a_0$ 
in the observed radial acceleration $\alpha = \sqrt{a_0a_N}$ \citep{mil84}, satisfying
\begin{eqnarray}
    a_0 = \frac{c^2}{2\pi}\sqrt{J}.
    \label{EQN_a0}
\end{eqnarray}
Direct estimates of $a_0$ derive from the bTFR relation \citep{mcc12}. 
Alternatively, $a_0$ is frequently estimated using interpolation
functions of rotation curves resolved over radius \citep{san90,lel16}. 
These two estimates are asymptotically consistent, provided the latter is corrected for systematic errors \citep{van18b}. 
Based on (\ref{EQN_a0}) with $q_0\simeq -1$ at present, we have \citep{van18} 
\begin{eqnarray}
    \frac{a_N}{\alpha} = \sqrt{2\pi}\left(1-q\right)^{-1/4}\zeta^{1/2}\simeq 2.1\,\zeta^{1/2}
    \label{EQN_a00}
\end{eqnarray}
at $z\simeq 0$. 
This result is in accurate agreement with SPARC \citep{van18}.
\mbox{}\\

\section{Conclusions and outlook} 

For the first time, we identify a sensitivity of weak gravitation below $a_{dS}$ across all redshifts, marked by a sharp 
$C^0$-transition across $a_{dS}$ to non-Newtonian motion
with asymptotic behavior parameterized by $a_0$ in $J$
(\ref{EQN_a0}) based on (\ref{EQN_N2d}-\ref{EQN_a00}).
The $C^0$-transition across the corresponding transition radius (\ref{EQN_rt}) bears out in SPARC \citep{van18} at $r_t\simeq r_c/\sqrt{2\pi}$, {\em inside} the previously proposed critical radius $r_c=\sqrt{R_G/a_0}c$ \citep{san90}. 

The non-Newtonian acceleration (\ref{EQN_N2d}) reveals short time scale orbital motion of the bTFR and fast gravitational collapse in early galaxy formation. The latter is accelerated by about an order of magnitude by $\sim a_0^{1/4}$ \citep{van24e}, i.e., earlier by a factor $\sim J^{1/8}={\cal O}\left(10^1\right)$ at cosmic dawn than what is expected in $\Lambda$CDM. 

We conclude that the sensitivity of weak gravitation to $J$ establishes a unified origin of anomalous galaxy dynamics across all redshifts -- a key signature of infrared gravitation. 
For low redshift galaxy surveys, it predicts for bTFR and galaxy dynamics that $a_0^\prime(0)<0$ due to $q_0^\prime(0)>0$ in (\ref{EQN_a0}) and a sharp $C^0$-transition across $a_{dS}$ distinct from $\Lambda$CDM galaxy models. For high redshift galaxy surveys, it accounts for early galaxy formation at cosmic dawn, preserving $\Lambda$CDM for early cosmology.

The transition (\ref{EQN_rt}) to anomalous acceleration below $a_{dS}$ obviates the need for CDM on galactic scales. 
Yet, a cosmological distribution of CDM is necessary to preserve a three-flat FLRW cosmology
(\ref{EQN_01}). This paradox points to ultra-light DM with de Broglie wavelength $ > r_t$, which sets a mass upper bound of 
about $m_Dc^2\lesssim 3\times 10^{-21}$eV of dark matter particles \citep{van24d}. This is consistent with the lower bound $m_Dc^2\gtrsim 10^{-22}$eV for wave-like dark matter ($\psi$DM) \citep{shi14,hui17,leu18,poz20,dem20,bro20,poz21,her19,poz23,bau22}, i.e., fuzzy dark matter \citep{hu00,pee00,sik09}).
Accordingly $r_t$ sets the smallest scale of the distribution of DM consistent with ALMA surveys \citep{ino23}.

 \begin{acknowledgements}
     We gratefully acknowledge the organizers of the Corfu2023 Workshop on {\em Tensions in Cosmology} where some of this work was presented \citep{van24b} and detailed comments on the present manuscript from M.A. Abchouyeh. This work is supported, in part, by the National Research Foundation of Korea, No. RS-2024-00334550.
\end{acknowledgements}

%\bibliographystyle{99}

%% This command is needed to show the entire author+affiliation list when
%% the collaboration and author truncation commands are used.  It has to
%% go at the end of the manuscript.
%\allauthors

\end{document}